\documentclass[letterpaper]{article}
\usepackage[utf8]{inputenc}

\usepackage{amsmath}
\usepackage{amsfonts}
\usepackage{amssymb}
\usepackage{amsthm}
\usepackage{authblk}
\usepackage{booktabs}
\usepackage{cite}
\usepackage{enumerate}
\usepackage{graphicx}
\usepackage[left=2cm,right=2cm,top=2cm,bottom=2cm]{geometry}
\usepackage{mathabx}
\usepackage{multirow}
\usepackage{subcaption}
\usepackage[normalem]{ulem}

\usepackage{color}

\newcommand{\diff}{\mathrm{d}}

\author{Pasquale Bosso\thanks{pasquale.bosso@uleth.ca}}
\affil{University of Lethbridge,\protect\\ 4401 University Drive, Lethbridge, Alberta, Canada, T1K 3M4\vspace{1em}}
\title{On the quasi-position representation in theories with a minimal length}
\date{}

\begin{document}

\maketitle

\begin{abstract}
Quantum mechanical models with a minimal length are often described by modifying the commutation relation between position and momentum.
Although this represents a small complication when described in momentum space, at least formally, the (quasi-)position representation acquires numerous issues, source of misunderstandings.
In this work, we review these issues, clarifying some of the aspects of minimal length models, with particular reference to the representation of the position operator.
\end{abstract}

\section{Introduction}

In phenomenological approaches to quantum gravity, it is often expected that a minimal measurable length has a fundamental role at high energies \cite{Garay1995_1,Amelino-Camelia2013}.
This expectation is motivated by various approaches to quantum gravity \cite{Amati1989_1,Gross1988_1,Rovelli1995}.
Furthermore, a minimal measurable length is expected in the shape of a minimal uncertainty in position when the same heuristic process of Heisenberg's microscope is considered including gravity or when it is applied to systems in which gravity is a fundamental component, such as black holes \cite{Mead:1964zz,Maggiore1993_1,Scardigli1999_1}.
All this suggested a modification of the uncertainty principle of Quantum Mechanics (QM) with the inclusion of a minimal length and led to the development of the Generalized Uncertainty Principle (GUP).
Such description of a minimal length has been elaborated in various forms:
as a modification of the uncertainty relation without the requirement of a particular representation for the corresponding quantum operators \cite{Scardigli1999_1,Scardigli2015,Casadio:2020rsj}; as a consequent modification of classical mechanics \cite{Mignemi2012,Pramanik2013_1,Pramanik2014_1,Chashchina:2019uyt,Bosso2018}; and as a modification of the position-momentum commutation relation \cite{Maggiore1993_2,Kempf1995_1,Ali2011_1,BossoPhD}.
In this work, we will focus on this last approach.
In particular, we will consider the case of a generic commutation relation in one dimension of the form
\begin{equation}
	[\hat{q},\hat{p}] = i \hbar f(\hat{p}).  \label{eqn:GUP_gen}
\end{equation}
Using \cite{Kempf1995_1} as a guideline, we will then elaborate on this model, obtaining constraints on the function $f$ and studying the various aspects implied by it.
Our focus here is to study and clarify aspects related with the position operator, especially in the quasi-position representation introduced in \cite{Kempf1995_1}.
Specifically, the position operator is often represented as a multiplicative operator.
Such approach is in contrast with the lack of position eigenstates implied by a minimal uncertainty in position.

It is worth emphasizing that here we will focus on the case a minimal measurable length is present, \emph{i.e.}, any state will not be able to be localized better than a quantity given by the model.
In fact, were this not the case, a set of position eigenstates would still exist, making the concept of quasi-position of little use.
Furthermore, as we will see below, the presence of a vanishing minimal uncertainty in position is in tension with the setting of this paper, given by Eq.\eqref{eqn:GUP_gen} and following \cite{Kempf1995_1}.

This paper is structured as follows:
In Section 2, we consider the case of a generic commutation relation between position and momentum, obtaining the maximally localized states, developing the corresponding integral transform, and studying the representation of position and momentum operators in quasi-position space.
In Section \ref{sec:second_degree}, we specialize the results to the case $f$ be a second degree polynomial in its argument.
This case is relevant since much of the literature is on this particular model with a specific choice of the parameters introduced below.
In Section \ref{sec:box}, we consider the case of a particle in a box as an instructive example in the framework of systems with a minimal length.
Furthermore, in Section \ref{sec:barrier}, we study the case of a potential barrier, with particular emphasis on the transmission coefficient for such system.
Finally, we conclude in Section \ref{sec:conclusions} by clarifying some aspects derived from our analysis.

\section{Generalization to arbitrary commutation relations}

In this section, we are going to follow the same arguments developed in \cite{Kempf1995_1} but applied to an arbitrary commutation relation Eq.\eqref{eqn:GUP_gen},
as long as the function $f(p)$ is sufficiently well-behaved.
Here, as well as in what follows, we will use the symbols $\hat{A}$ and $A$ for the operator and the c-number associated with the quantity $A$, respectively.
One first aspect to notice is that, since the commutator of two observables is anti-Hermitian, the function $f$, when regarded as a function of a real variable, has real values.

Let us start with the momentum representation of the position and momentum operators compatible with the commutator in Eq.\eqref{eqn:GUP_gen},
\begin{align}
	\hat{q} = & i \hbar f(p) \frac{\diff}{\diff p}, & \hat{p} = & p.
\end{align}
It is worth noticing that different representations for these two operators compatible with Eq.\eqref{eqn:GUP_gen} may be allowed \cite{Bishop2020}.
However, in what follows, we will adhere to the ideas introduced in \cite{Kempf1995_1}.
Thus, a maximally localized state, if it exists, is a solution of the following differential equation
\begin{equation}
	\frac{\diff}{\diff p} \psi_{\langle q \rangle} (p) = - \frac{1}{f(p)} \left\{\frac{i}{\hbar} \langle q \rangle + \frac{\langle f(p) \rangle (p - \langle p \rangle)}{2 (\Delta p)^2}\right\} \psi_{\langle q \rangle}(p), \label{eqn:min_unc_gen}
\end{equation}
where the subscript ${\langle q \rangle}$ indicates that the wave function $\psi_{\langle q \rangle}(p)$ corresponds to a particular value of the position expectation value, considered as a parameter.
Furthermore, the expectation values in the previous equation are computed with respect to $\psi_{\langle q \rangle}(p)$.
In what follows, it is convenient introducing the auxiliary momentum $p_0$ defined as the momentum conjugate to $\hat{q}$.
Since with respect to this new variable, the position operator is $\hat{q} = i \hbar \frac{\diff}{\diff p_0}$, it is easy to see that $p_0$ is related to the physical momentum through
\begin{equation}
	\frac{\diff p}{\diff p_0} = f(p) \qquad \text{or} \qquad p_0 = p_0(p) = \int \frac{\diff p}{f (p)}. \label{eqn:p_to_p_0}
\end{equation}
A similar relation has been obtained in \cite{Hossenfelder2003}.
Notice that in principle $p_0$ acquires values on a subset of $\mathbb{R}$.
In particular, we will consider the case in which the function $p_0 = p_0(p) : \mathbb{R} \rightarrow (a,b) \subset \mathbb{R}$ is invertible in the same set $(a,b)$.
Furthermore, imposing that $p_0 \simeq p$ for small values of $|p|$ requires $a<0<b$.
For $p_0(p)$ to be invertible, this function has to be monotonic, thus $f(p)$ cannot change sign.
Therefore, in particular it cannot be an odd function.
Moreover, since in standard QM $f(p)=1$, the function $f(p)$ has to be non-negative so to have the correct limit for low momenta.
A further motivation for this property is related with the meaning of the function $f(p)$ itself in the context of GUP, as described in \cite{Kempf1995_1}.
In fact, since such a function is related with the measure of momentum space, it cannot be negative.
To avoid $f(p)$ from changing sign, one can simply restrict the domain for the physical momentum variable $p$ to a subset of $\mathbb{R}$ containing 0 and in which $f(p)$ does not change sign.
This is the case of several works in the literature, \emph{e.g.} \cite{Jizba:2009qf,Ong:2018zqn,Buoninfante:2019fwr,Bosso:2020ztk}.
However, for simplicity, in what follows we will continue considering that $p$ acquires any real value unless otherwise specified.

We then see that, although it is necessary to impose a particular measure in momentum space so that the position operator is symmetric when represented in terms of the variable $p$, when momentum space is expressed in terms of $p_0$ the measure is the usual one.
Moreover, it is worth mentioning that in general $-a \neq b$ when imposing $p_0(0)=0$.
The equality is fulfilled only when $f(p)$ is an even function.
Below, we will in fact see an example of a non-even function $f(p)$ leading to a non-symmetric interval $(a,b)$ with respect to the value $p_0=0$.

As for Eq.\eqref{eqn:min_unc_gen}, we can write its solutions in the form
\begin{equation}
	\psi_{\langle q \rangle}(p) = \chi(p) \exp\left[- i \frac{\langle q \rangle \, p_0(p)}{\hbar}\right], \label{eqn:min_unc_exp}
\end{equation}
where $\chi(p)$ is a function that depends only on $p$.
In fact, in this way Eq.\eqref{eqn:min_unc_gen} reduces to
\begin{equation}
	\frac{\diff}{\diff p} \chi (p) = - \frac{\langle f(\hat{p}) \rangle (p - \langle p \rangle)}{2 f(p) (\Delta p)^2} \chi(p).
\end{equation}
Symbolically, the solution for this equation is
\begin{equation}
	\chi(p) = \exp \left[- \frac{\langle f(\hat{p}) \rangle}{2 (\Delta p)^2} \left(\int \diff p \frac{p - \langle p \rangle}{f(p)}\right)\right],
\end{equation}
or, in terms of the quantity $p_0$,
\begin{equation}
	\chi(p_0) = \exp\left[- \frac{\langle f(\hat{p}) \rangle}{2 (\Delta p)^2} \left(\int \diff p_0 p(p_0) - \langle p \rangle p_0\right)\right].
\end{equation}
It is worth noticing that in the case of standard QM, one obtains the usual Gaussian function.
Furthermore, notice that $\chi(p) \in \mathbb{R}$.
In cases of QM with a minimal length, the maximally localized states are obtained when the values of the various parameters, namely $\langle p \rangle$, $\Delta p$, $\langle f(p)\rangle$, are such that the uncertainty in position is the smallest possible compatibly with the uncertainty relation derived from Eq.\eqref{eqn:GUP_gen}.
In what follows, we will thus assume that such condition is fulfilled and, therefore, that the parameters above have given values.

Following \cite{Kempf1995_1}, it is possible to use the complex conjugate version of the function thus found as kernel for an integral transform.
The corresponding new space is what has been called ``quasi-position space'' of variable $\xi \equiv \langle q \rangle$.
That is, it is possible to introduce the following two transformations
\begin{subequations}
	\begin{align}
		\mathcal{T}^{-1}[\phi](\xi) = & \frac{1}{\sqrt{2 \pi \hbar}} \int_{-\infty}^{\infty} \frac{\diff p}{f(p)} \psi^\star (p,\xi) \phi(p),\label{eqn:anti-transform}\\
		\mathcal{T}[\phi](p) = & \frac{1}{\sqrt{2 \pi \hbar}} \int_{-\infty}^{\infty} \diff \xi \left[\psi^\star (p,\xi)\right]^{-1} \phi(\xi), \label{eqn:transform}
	\end{align} \label{eqn:transforms}
\end{subequations}
where $\psi (p,\xi) = \psi_\xi (p)$ is regarded as a function of both $p$ and $\xi$.
We used the symbols $\mathcal{T}$ and $\mathcal{T}^{-1}$ because they are in fact one the inverse of the other, since
\begin{multline}
	\mathcal{T}^{-1} \left[\mathcal{T}[\phi]\right](\xi')
	= \frac{1}{2 \pi \hbar} \int_{-\infty}^{\infty} \frac{\diff p}{f(p)} \psi_\xi^\star (p,\xi') \int_{-\infty}^{\infty} \diff \xi \left[\psi_\xi^\star (p,\xi)\right]^{-1} \phi(\xi)
	= \frac{1}{2 \pi \hbar}  \int_{-\infty}^{\infty} \diff \xi \int_{a}^{b} \diff p_0 \exp\left[- i \frac{(\xi - \xi') p_0}{\hbar}\right] \phi(\xi) \\
	= \frac{i}{2 \pi}  \int_{-\infty}^{\infty} \diff \xi \frac{1}{\xi - \xi'} \left\{\exp\left[- i \frac{(\xi - \xi') b}{\hbar}\right] - \exp\left[- i \frac{(\xi - \xi') a}{\hbar}\right]\right\}\phi(\xi) = \phi (\xi').
\end{multline}
Furthermore, it is easy to find the following relations concerning the convolution and cross-correlation of two functions, respectively,
\begin{align}
	\mathcal{T}[\phi_1] (p) \, \mathcal{T}[\phi_2] (p)
	= \frac{1}{\sqrt{2 \pi \hbar} \chi(p)}  \mathcal{T}\left[\int_{-\infty}^{\infty} \diff \xi \phi_1(\xi) \phi_2(\Xi - \xi)\right] (p)
	= & \frac{1}{\sqrt{2 \pi \hbar} \chi(p)} \mathcal{T}[\phi_1 * \phi_2] (p),\\
	\mathcal{T}[\phi_1]^\star (p) \, \mathcal{T}[\phi_2] (p)
	= \frac{1}{\sqrt{2 \pi \hbar} \chi(p)}  \mathcal{T}\left[\int_{-\infty}^{\infty} \diff \xi \phi_1^\star(\xi) \phi_2(\Xi + \xi)\right] (p)
	= & \frac{1}{\sqrt{2 \pi \hbar} \chi(p)} \mathcal{T}\left[\phi_1 \star \phi_2\right] (p). \label{eqn:cross-correlation_property_GUP}
\end{align}
These relations are similar to those obtained for the ordinary Fourier transform.
However, in the present case, these relations hold only for the transform $\mathcal{T}$ and not for the anti-transform $\mathcal{T}^{-1}$.
Such feature is due to the presence of the functions $f(p)$ and $\chi(p)$ in Eq.\eqref{eqn:anti-transform}.
In fact, since both such functions depend on $p$, the product $\mathcal{T}^{-1}[\phi_1] (p) \, \mathcal{T}^{-1}[\phi_2] (p)$ cannot be cast in the form of Eq.\eqref{eqn:anti-transform} because both $f(p)$ and $\chi(p)$ belonging to each of the anti-transforms remain trapped in the respective integrals.

Using the integral transform defined above, it is possible to find the generic momentum eigenfunction in quasi-position space, \emph{i.e.}
\begin{equation}
	\phi_{\tilde{p}}(\xi) 
	= \mathcal{T}^{-1}[\delta(\tilde{p} - p)](\xi)
	= \frac{1}{\sqrt{2 \pi \hbar}} \frac{\chi(\tilde{p})}{f(\tilde{p})} \exp \left[i \frac{\xi \, p_0(\tilde{p})}{\hbar}\right]~. \label{eqn:momentum_eigen_position}
\end{equation}
Notice that it is a plane wave of wave number $k = p_0 / \hbar$.
In this sense, de Broglie relation is modified in models with modified uncertainty relations.
In a different sense, one can use this relation to define the auxiliary momentum $p_0$.
Furthermore, this is clearly the wave function of a free particle.
In fact, since the Hamiltonian of a free particle is just proportional to $\hat{p}^2$, the Hamiltonian and momentum operators share the same set of eigenfunctions.
Specifically, any value of the energy will correspond to a pair of solutions for the free-particle Schr\"odinger equation, each of which of the form in Eq.\eqref{eqn:momentum_eigen_position} and opposite eigenvalues $\tilde{p}$, simply representing left- and right-moving waves.
However, it is worth observing that, when $f(p)$ is not an even function, and thus when $p_0(p)$ is not an odd function, the two solutions do not have opposite values of $p_0(\tilde{p})$ and therefore they will not have opposite wave numbers.
This aspect will have important implications in the two examples shown below.

Using again the transform in Eq.\eqref{eqn:anti-transform}, the representation of the position operator in the new space is
\begin{equation}
	\hat{q} = \xi + i \hbar \frac{\langle f(\hat{p}) \rangle}{2 (\Delta p)^2}  (\hat{p} - \langle p \rangle). \label{eqn:q_op_gen}
\end{equation}
As for the momentum operator, Eq.\eqref{eqn:min_unc_exp} is enough to say that the quasi-position representation of the momentum operator is $\hat{p} = p(\hat{p}_0)$, with $p(p_0)$ the inverse function of $p_0(p)$ and
\begin{equation}
	\hat{p}_0 = - i \hbar \frac{\diff}{\diff \xi}.
\end{equation}
One can then easily see that this is consistent with the commutation relation we started with since
\begin{equation}
	[\hat{q},\hat{p}] = i \hbar \frac{\diff \hat{p}}{\diff \hat{p}_0} = i \hbar f(\hat{p}).
\end{equation}
Notice that the kernel in the integral transform is a function of the minimal uncertainty product according to Eq.\eqref{eqn:min_unc_gen}.
Thus, in all generality, even when the kernel is not a maximally localized function, the momentum operator in the corresponding new space is represented by
\begin{equation}
	\hat{p} = p\left(- i \hbar \frac{\diff}{\diff \xi}\right),
\end{equation}
with $p(p_0)$ the inverse function of $p_0(p)$ found in Eq.\eqref{eqn:p_to_p_0}.
We highlight the fact that this is always valid as long as the kernel of the integral transform has minimal uncertainty product.

To conclude this analysis, it is worth paying attention to the second term on the right hand side of Eq.\eqref{eqn:q_op_gen}.
First, we notice that it does not contribute to the commutation relation since it is a linear function of the operator $\hat{p}$ alone.
In fact, all the other terms are constant and depend on the particular minimal uncertainty state chosen as kernel for the integral transforms in Eqs.\eqref{eqn:transforms}.
Second, we notice that it poses an apparent problem because of the factor $i \hbar$.
In fact, it may seem that the position operator is no longer Hermitian.
However, as argued in \cite{Kempf1995_1}, since the functions $\psi_{\langle q \rangle}(p)$ for different choices of $\langle q \rangle$ are in principle not orthogonal, the scalar product of two functions in quasi-position space has to be taken necessarily after having transformed the two functions to momentum space.
That is, given two functions $\phi_1(\xi)$ and $\phi_2(\xi)$, their scalar product is
\begin{equation}
	\langle \phi_1 | \phi_2 \rangle
	= \int_{-\infty}^{\infty} \frac{\diff p}{f (p)} \mathcal{T}[\phi_1]^\star (p) \mathcal{T}[\phi_2] (p)
	= \frac{1}{\sqrt{2 \pi \hbar}} \int_{-\infty}^{\infty} \frac{\diff p}{f (p) \chi(p)} \mathcal{T} \left[\phi_1 \star \phi_2\right] (p).
\end{equation}
Clearly, since the scalar product does not depend on the particular representation, the operator $\hat{q}$ has to be symmetric in both representations, despite the presence of the imaginary unit in Eq.\eqref{eqn:q_op_gen}.
In other words, the presence of the imaginary unit and of an imaginary constant in the position operator, namely the term $i \hbar \langle f(\hat{p}) \rangle \langle p \rangle / 2 (\Delta p)^2$, is a by-product of the particular representation in use.
This makes the variable $\xi$ hardly interpretable as a position coordinate.
Furthermore, it is worth noticing that a hypothetical operator $\hat{\xi}$ acting multiplicatively in quasi-position space is not Hermitian.
In fact, using the transformation in Eq.\eqref{eqn:transform} or simply inspecting Eq.\eqref{eqn:q_op_gen} above, we have in momentum space
\begin{equation}
	\hat{\xi} = \hat{q} - i \hbar \frac{\langle f(\hat{p}) \rangle}{2 (\Delta p)^2}  (\hat{p} - \langle p \rangle),
\end{equation}
which is evidently non-Hermitian.

Another argument to show that an operator acting by multiplying a wave function by the quasi-position coordinate $\xi$ is not physical, and in particular that it does not represent a position coordinate, is the following.
First, let us notice that the factor $\langle f(\hat{p}) \rangle$ is the expectation value of the function of operator $f(\hat{p})$ on the minimal uncertainty states.
Even assuming that the expectation values of powers of momentum with respect such states are small with respect to the corresponding power of a characteristic momentum, we will still obtain a non-vanishing factor.
Thus, in this ``low-momentum'' limit, in which $\langle f(\hat{p})\rangle \rightarrow 1$, the position operator would be written as
\begin{equation}
	\hat{q} = \xi + \frac{i \hbar}{2 (\Delta p)^2}  (\hat{p} - \langle p \rangle).
\end{equation}
This is in fact what one would obtain defining an integral transform from momentum to quasi-position spaces using minimal uncertainty states of standard QM, represented in momentum space by
\begin{equation}
	\psi_\xi (p) \propto \exp\left[- \frac{(p - \langle p \rangle)^2}{(2 \Delta p)^2}\right] \exp \left[- i \frac{\xi p}{\hbar}\right],
\end{equation}
as kernel functions.
Since the variable $\xi$ defined through such transform cannot be considered as the position coordinate of standard QM, it cannot be considered a position coordinate in any other model.
Although this inconvenience is explained in QM as a poor choice of the transform, in models with a minimal length it is a fundamental issue, although consistent with the problem of introducing a minimal uncertainty in position.
In fact, since in this case a proper position-space description is not possible due to the lack of position eigenstates, we are forced to resort to maximally localized states and the corresponding quasi-position space with the consequent issues involving the position operator.

\section{Second degree commutation relation} \label{sec:second_degree}

In what follows, as a particular example, we will consider the following generalized commutation relation for a one-dimensional system
\begin{equation}
	[\hat{q},\hat{p}] = i \hbar \left[1 - 2 \delta \hat{p} + (\delta^2 + \epsilon) \hat{p}^2\right]~, \label{eqn:GUP}
\end{equation}
with
\begin{align}
	\delta = & \frac{\delta_0}{M_\mathrm{Pl} c}, &
	\epsilon = &
	\frac{\epsilon_0}{(M_\mathrm{Pl} c)^2},
\end{align}
and $M_\mathrm{Pl}$ and $c$ being the Planck mass and the speed of light in vacuum, respectively.
Here, $\delta_0$ and $\epsilon_0$ are two dimensionless parameters of order 1 that determine the particular GUP model, (\emph{e.g.}, the model in \cite{Kempf1995_1} is given by $\delta_0 = 0$ and $\epsilon_0 = 1$, while the model in \cite{Ali2011_1} is given by $\delta_0 = 1, \epsilon_0 = 3$).
As we will see, they are related to the expectation value and uncertainty squared of momentum for a state of minimal uncertainty in position.
This model is consistent with the discussion above when $\delta_0 \in \mathbb{R}$ and $\epsilon_0 > 0$.
Furthermore, it represents the most general one-dimensional case up to second order in the inverse Planck momentum.
Therefore, any result obtained for a generic function $f(\hat{p})$ in the previous section, when expanded in series up to second order in the inverse Planck momentum, has to agree with what we are going to show.

As for the auxiliary momentum, we then have
\begin{equation}
	p_0(p) = \frac{1}{\sqrt{\epsilon}} \arctan\left[\frac{- \delta + (\delta^2 + \epsilon) p}{\sqrt{\epsilon}}\right] + \frac{1}{\sqrt{\epsilon}} \arctan\left(\frac{\delta}{\sqrt{\epsilon}}\right), \label{eqn:p_to_p_0_2}
\end{equation}
where the arbitrary constant in the definition of $p_0$ has been chosen so that $p_0(0) = 0$.
The function $p_0(p)$ has values in the interval
\begin{equation}
	p_0(p) \in \left]\frac{1}{\sqrt{\epsilon}} \arctan\left(\frac{\delta}{\sqrt{\epsilon}}\right) - \frac{\pi}{2 \sqrt{\epsilon}} \quad, \quad \frac{1}{\sqrt{\epsilon}} \arctan\left(\frac{\delta}{\sqrt{\epsilon}}\right) + \frac{\pi}{2 \sqrt{\epsilon}}\right[. \label{eqn:interval_p_0}
\end{equation}
It is worth noticing that the arbitrary constant is proportional to the Planck momentum and that it vanishes in models with $\delta_0 = 0$.
Furthermore, the same constant represents the centre of the interval of values of $p_0$.
Thus, the parameter $\delta_0$ shifts the centre, and therefore the entire interval, with respect to the value $p_0=0$.
This is simply an effect of the anisotropy of this model characterized by the linear term in Eq.\eqref{eqn:GUP}.
In fact, as pointed out in the previous section, any non-even function $f(\hat{p})$ will lead to such feature.

\begin{figure}
	\centering
	\includegraphics[width=0.75\textwidth]{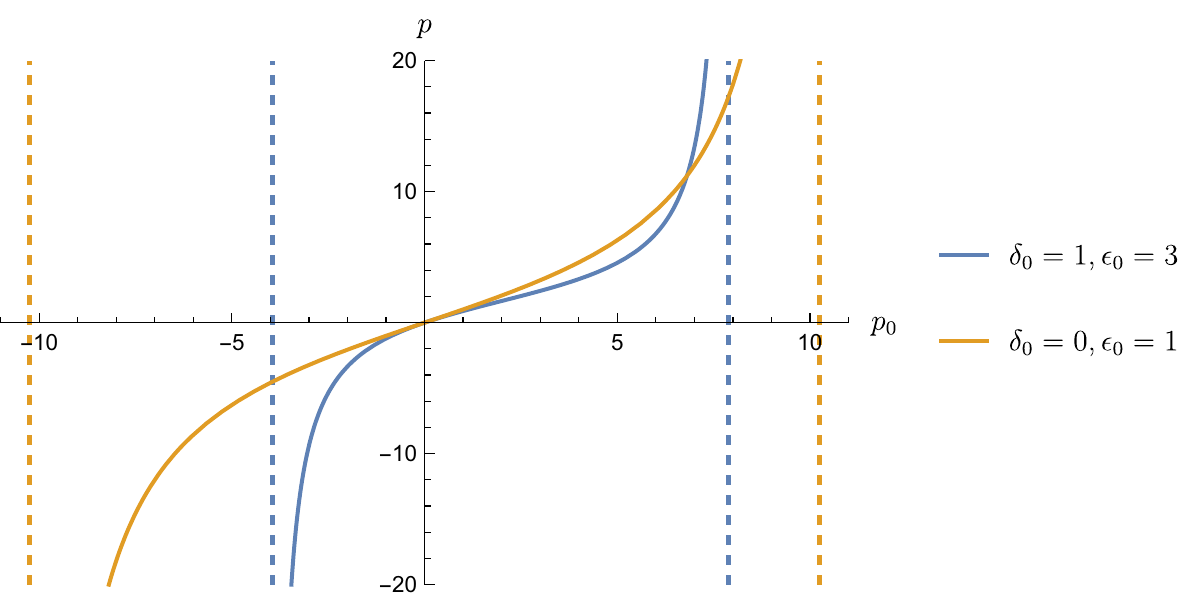}
	\caption{Plot of the function $p(p_0)$ for two different choices of the parameters.
	Both quantities, $p$ and $p_0$, are reported in SI units.
	The pair of dashed lines identify the asymptotes for the two models.
	Notice that both solid lines pass through the point $p_0=0,p=0$, although the blue solid line is not symmetric with respect to the line $p_0=0$.} \label{fig:pofp_0}
\end{figure}
In Fig.\ref{fig:pofp_0}, we plotted the function $p(p_0)$, inverse of Eq.\eqref{eqn:p_to_p_0_2} for the choices of the parameters $\delta_0$ and $\epsilon_0$ corresponding to \cite{Kempf1995_1} and \cite{Ali2011_1}.
It is worth noticing that the domain of the function $p(p_0)$ is limited to the interval in Eq.\eqref{eqn:interval_p_0} and that, for the case of \cite{Ali2011_1}, the function is not symmetric with respect to $p_0=0$, as mentioned above.

Compatibly with the results in \cite{Bosso2018} and of the previous section, we define the position and momentum operators in momentum space as
\begin{align}
	\hat{q} = & i \hbar \left[1 - 2 \delta p + (\delta^2 + \epsilon) p^2\right] \frac{\diff}{\diff p}~, & \hat{p} = & p~. \label{eqn:ops_momentum}
\end{align}
We see, in fact, that the operators in Eq.\eqref{eqn:ops_momentum} fulfill the commutation relation in Eq.\eqref{eqn:GUP}, corresponding to the following uncertainty relation
\begin{equation}
	\Delta q \Delta p \geq \frac{\hbar}{2} \left\{1 - 2 \delta \langle p \rangle + (\delta^2 + \epsilon) [(\Delta p)^2 + \langle p \rangle^2] \right\}. \label{eqn:uncertainty_relation}
\end{equation}
The minimal position uncertainty compatible with this model, $\Delta q_\text{min} = \hbar \sqrt{\epsilon}$, is obtained for a state such that
\begin{align}
	\langle p \rangle = & \frac{\delta}{\delta^2 + \epsilon},&
	\Delta p = & \frac{\sqrt{\epsilon}}{\delta^2 + \epsilon},&
	\langle p^2 \rangle = & (\Delta p)^2 + \langle p \rangle^2 = \frac{1}{\delta^2 + \epsilon}. \label{eqn:p_unc_min}
\end{align}
It is interesting to observe that, differently from the results in \cite{Kempf1995_1}, the minimal uncertainty in position is obtained for a non-vanishing expectation value of momentum when a linear term is present in Eq.\eqref{eqn:GUP}. 
Furthermore, it only depends on the parameters of the model considered here and is of the order of the Planck momentum.

With these values for the expectation values and uncertainties, the solution of Eq.\eqref{eqn:min_unc_gen} is
\begin{equation}
	\psi_{\langle q \rangle} = \chi(p) \exp\left[- i \frac{\langle q \rangle \,\, p_0(p)}{\hbar}\right], \qquad \text{with} \quad 
	\chi(p) = \frac{1}{\sqrt{1 - 2 \delta p + (\delta^2 + \epsilon) p^2}}.
\end{equation}
This wave function represents a state of minimal uncertainty in position whose position expectation value is $\langle q \rangle$.
On the other hand, treating the quantity $\xi \equiv \langle q \rangle$ as a variable, we can use this function as a kernel for an integral transform compatible with GUP from the momentum space to the space of functions of variable $\xi$.
In fact, let us define two transforms, \emph{i.e.}, one from $p$-space to $\xi$-space, and another from $\xi$-space to $p$-space
\begin{subequations}
	\begin{align}
	\mathcal{T} \left[\phi\right] (p) = & \frac{1}{\sqrt{2 \pi \hbar}} \int_{-\infty}^\infty \diff \xi ~ \sqrt{1 - 2 \delta p + (\delta^2 + \epsilon) p^2} \exp \left(- i \frac{\xi p_0(p)}{\hbar}\right) \phi (\xi), \label{eqn:fourier} \\
	\mathcal{T}^{-1} \left[\phi\right] (\xi) = & \frac{1}{\sqrt{2 \pi \hbar}} \int_{-\infty}^\infty \frac{\diff p}{[1 - 2 \delta p + (\delta^2 + \epsilon) p^2]^{3/2}} \exp \left(i \frac{\xi p_0(p)}{\hbar}\right) \phi (p). \label{eqn:anti-fourier}
	\end{align}
\end{subequations}
These two transforms correspond to those in Eqs.\eqref{eqn:transforms}.
Thus, we have
\begin{equation}
	\mathcal{T}^{-1} \left[\mathcal{T} \left[\phi\right]\right] (\xi) = \phi (\xi). \label{eqn:invertibility}
\end{equation}
Furthermore, they produce the following function for a free particle
\begin{equation}
	\psi_{\tilde{p}}(\xi) 
	= \mathcal{T}^{-1}[\delta(\tilde{p} - p)](\xi)
	= \frac{1}{\sqrt{2 \pi \hbar}} \frac{1}{[1 - 2 \delta \tilde{p} + (\delta^2 + \epsilon) \tilde{p}^2]^{3/2}} \exp \left[i \frac{\xi \, p_0(\tilde{p})}{\hbar}\right]~.
\end{equation}

In quasi-position representation, and with the values in Eq.\eqref{eqn:p_unc_min}, we have
\begin{align}
	\hat{q} = & \xi + i \hbar \left[(\delta^2 + \epsilon) \hat{p} - \delta\right], &
	\hat{p} = & \frac{\sqrt{\epsilon} \tan \left[\sqrt{\epsilon} \hat{p}_0 - \arctan \left(\frac{\delta}{\sqrt{\epsilon}}\right)\right] + \delta}{\delta^2 + \epsilon}, &
	\hat{p}_0 = & - i \hbar \frac{\diff}{\diff \xi}.
\end{align}

Finally, it is easy to see that both position and momentum operators have the correct limit for $1/M_\text{Pl} c \rightarrow 0$.
Furthermore, they reproduce what found in \cite{Kempf1995_1} for $\delta_0 = 0$ and $\epsilon_0 = 1$.
Moreover, up to second order in $1/M_\text{Pl} c$, the momentum operator has the following form, used \emph{e.g.} in \cite{Ali2011_1}
\begin{equation}
	\hat{p} = \hat{p}_0 \left[ 1 - \delta \hat{p}_0 + \left(\delta^2 + \frac{\epsilon}{3}\right) \hat{p}_0^2 \right]. \label{eqn:p_of_p_0_2}
\end{equation}
However, notice that, up to the same order in the inverse Planck momentum, the position operator in quasi-position space is
\begin{equation}
	\hat{q} = \xi + i \hbar [(\delta^2 + \epsilon) \hat{p}_0 - \delta].
\end{equation}
This expression shows the difficulties rising in the use of quasi-position representation with GUP pointed out at the end of the previous section, even at an approximate level.
In fact, one is usually tempted to modify the representation of the momentum operator using expressions similar to Eq.\eqref{eqn:p_of_p_0_2}, while retaining the position operator as a multiplicative one or modifying it using expressions like
\begin{equation}
	\hat{q} = x \left[1 + \mathcal{P}(\hat{p}_0)\right],
\end{equation}
where $x$ is a real variable and $\mathcal{P}$ is a polynomial of given degree and zero constant term.
If the coefficients of the polynomial $\mathcal{P}$ are real, the operator $\hat{q}$ will necessarily be different from the one obtained above.
Furthermore, based on the discussion of the previous section, the quantity $x$ is not associated with an actual position coordinate.

Based on the analyses in the previous section, the model in Eq.\eqref{eqn:GUP} requires some attention when $\epsilon_0 < 0$.
Specifically, with such a choice of the parameters we have intervals of $p$ in which $f(p)<0$.
In this case, considering the positive value of $\sqrt{-\epsilon_0}$, one is forced to restrict the domain for the variable $p$ to subsets of $\mathbb{R}$ on which the function $f(p)$ is non-negative.
Thus, momentum space is limited to the intervals
\begin{equation}
	p \in \left\{\begin{array}{clc}
		\displaystyle{\left(-\infty, \frac{\delta - \sqrt{-\epsilon}}{\delta^2 + \epsilon}\right)} & \text{for} \quad & \delta_0 > \sqrt{-\epsilon_0}\\
		[1.5em] \displaystyle{\left(\frac{\delta + \sqrt{-\epsilon}}{\delta^2 + \epsilon} , \frac{\delta - \sqrt{-\epsilon}}{\delta^2 + \epsilon}\right)} & \text{for} \quad & |\delta_0| < \sqrt{-\epsilon_0}\\
		[1.5em] \displaystyle{\left(\frac{\delta + \sqrt{\epsilon}}{\delta^2 + \epsilon} , + \infty\right)} & \text{for} \quad & \delta_0 < - \sqrt{- \epsilon_0}.
	\end{array}\right. \label{eqn:limited_p}
\end{equation}
It is easy and interesting noticing that, using the relations in Eq.\eqref{eqn:p_to_p_0}, while in the case $|\delta_0| > \sqrt{-\epsilon_0}$ both $p$ and $p_0$ are bounded (one from below, the other from above, or viceversa), in the case $|\delta_0| < \sqrt{-\epsilon_0}$, $p_0$ can acquire any real value while $p$ is limited on the interval described above.

\section{Particle in a box} \label{sec:box}

To elaborate further on the concept of position in these models, let us examine the case of a particle in a box.
An immediate complication arises as to how to define the boundaries of the box.
In fact, not being able to define sharp positions of points in the framework of this paper, we need to resort to alternative definitions.
We will base our argument on the following statement, a consequence of Eqs.\eqref{eqn:anti-transform} and \eqref{eqn:anti-fourier}: any wave function in quasi-position space can be regarded as a superposition of maximally localized states, each of position expectation value $\xi$.
Furthermore, we expect that any classical measurement of a physical quantity corresponds to evaluating the expectation values of the related quantum version.
Since considering a particle in a box means, from a classical point of view, that any position measurement in principle would return any value in the allowed region, the expectation value of the position is in the classically allowed region, which in turn, when working with maximally localized states, means that the quantity $\xi$ is in that same region.

Thus, let us consider a one-dimensional box of width $L$ such that $\xi \in ]0,L[.$
In this region, a particle in an energy eigenstate is described by a wave function $\psi(\xi)$ which is a superposition of free-particle wave functions as in Eq.\eqref{eqn:momentum_eigen_position}
\begin{equation}
	\psi(\xi) = A_n \frac{\chi(p_n)}{f(p_n)} \exp \left[i \frac{\xi \, p_0(p_n)}{\hbar}\right] + B_n \frac{\chi(-p_n)}{f(-p_n)} \exp \left[i \frac{\xi \, p_0(-p_n)}{\hbar}\right],
\end{equation}
with $A_n$ and $B_n$ constants.
Furthermore, by the argument above, we need to impose the following boundary conditions
\begin{align}
	\psi(\xi=0) = & 0, &
	\psi(\xi=L) = & 0.
\end{align}
The first boundary condition implies that the amplitudes $A_n$ and $B_n$ are related by
\begin{equation}
	B_n = - A_n \frac{\chi(p_n)}{\chi(-p_n)} \frac{f(-p_n)}{f(p_n)},
\end{equation}
thus obtaining
\begin{equation}
	\psi(\xi) = A_n \frac{\chi(p_n)}{f(p_n)} \left\{\exp \left[i \frac{\xi \, p_0(p_n)}{\hbar}\right] - \exp \left[i \frac{\xi \, p_0(-p_n)}{\hbar}\right]\right\}.
\end{equation}
As for the second boundary condition, it is satisfied when
\begin{equation}
	[p_0(p_n) - p_0(-p_n)] = \frac{2 \pi n \hbar}{L}, \qquad \text{with} \quad n \in \mathbb{N}. \label{eqn:wavenumber_box}
\end{equation}
When $f(p)$ is an even function, that is, when $p_0(p)$ is an odd function, \emph{e.g.} when $\delta_0=0$ in Eq.\eqref{eqn:GUP}, we have
\begin{equation}
	p_0(p_n) = \frac{n \pi \hbar}{L}, \qquad \text{with} \quad n \in \mathbb{N}.
\end{equation}
It is interesting to notice that this is the same relation that one would obtain in standard QM, but in terms of the auxiliary momentum $p_0$.
However, in general we may find left-moving and right-moving functions corresponding to different wavelengths due to a non-even dispersion relation $p^2(p_0) = 2 m E$.
Nonetheless, notice that $p_0$ may acquire values on a limited interval, as in the case of Eq.\eqref{eqn:interval_p_0}.
In fact, in this case we will have a maximum value for $n$ given by
\begin{equation}
	n_{\text{max}} = \left\lfloor \frac{L}{2 \hbar \sqrt{\epsilon}}\right\rfloor.
\end{equation}
It is worth noticing that such maximum integer does not depend on $\delta$.
The reason is that the range of the quantity $p_0$, as given in Eq.\eqref{eqn:interval_p_0}, does not depend on $\delta$.
Furthermore, notice that such value is of the order of $n_\text{max} \sim L / \ell_\text{Pl}$, with $\ell_\text{Pl}$ the Planck length.
This implies that some values of $L$ do not admit the presence of a particle in the box.
This happens for
\begin{equation}
	L < 2 \ell_\text{Pl} \sqrt{\epsilon_0}. \label{eqn:threshold}
\end{equation}
However, it is worth emphasizing that any value $L$ for the size of the box is in principle allowed and the size of the box does not present any form of quantization.

\begin{figure}
	\centering
	\begin{subfigure}{0.49\textwidth}
		\centering
		\includegraphics[width=\textwidth]{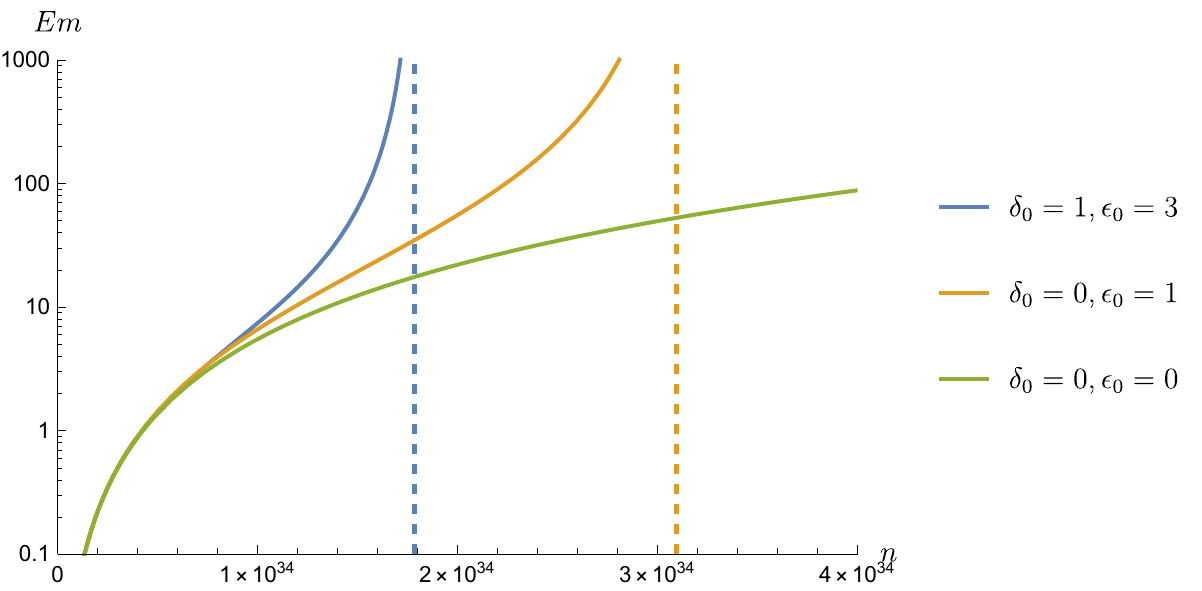}
		\caption{}
		\label{subfig:box1m}
	\end{subfigure}
	\begin{subfigure}{0.49\textwidth}
		\centering
		\includegraphics[width=\textwidth]{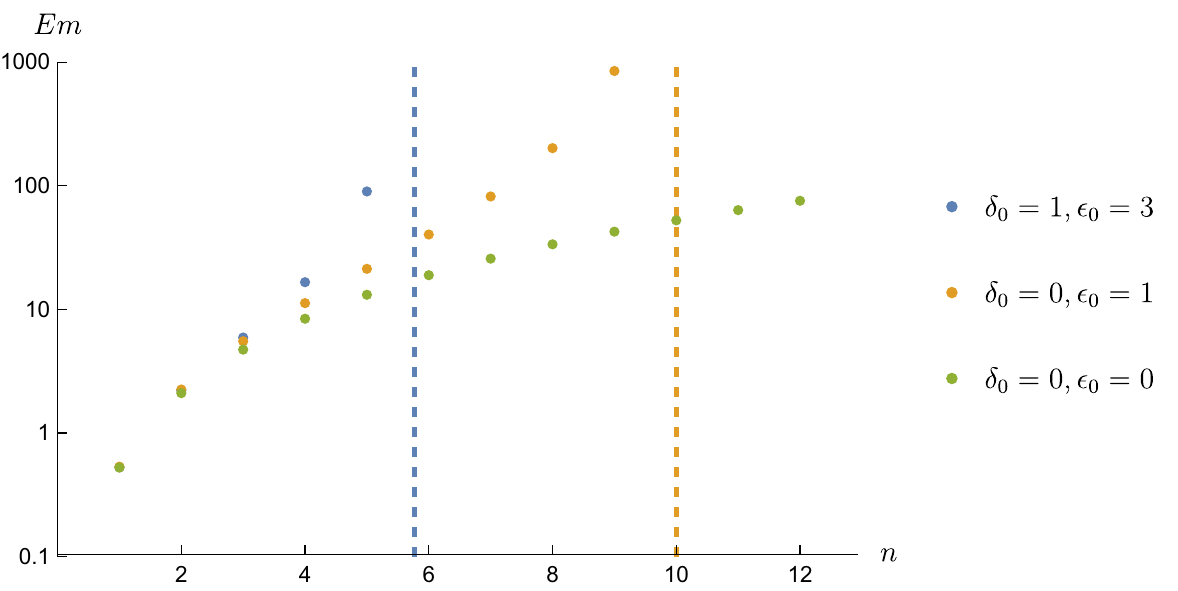}
		\caption{}
		\label{subfig:box20l}
	\end{subfigure}
	\caption{Energy times mass for a particle in a one-dimensional box of size $L=1$ m (Fig.\ref{subfig:box1m}) and for a box of size $L=20\ell_\text{Pl}$ (Fig.\ref{subfig:box20l}) as functions of the integer $n$.
	The $y$-axes are in logarithmic scale and the values are in SI units.
	The solid blue line (circles) is for the model in \cite{Ali2011_1}, the solid orange line (circles) is for \cite{Kempf1995_1}, the solid green line (circles) is for the standard case.
	The vertical dashed lines correspond to the maximum values for the integer $n$ for the two models.}
	\label{fig:box}
\end{figure}

Inverting Eq.\eqref{eqn:wavenumber_box} and specializing the function $f(p)$ to the case of Section \ref{sec:second_degree}, we have
\begin{equation}
	p_n
	= \frac{\sqrt{\delta^2 \sin^2 \left[\frac{2 \pi n \hbar \sqrt{\epsilon}}{L}\right] + \epsilon} - \sqrt{\epsilon} \cos \left[\frac{2 \pi n \hbar \sqrt{\epsilon}}{L}\right]}{\left(\delta^2 + \epsilon\right) \sin \left[\frac{2 \pi n \hbar \sqrt{\epsilon}}{L}\right]}.
\end{equation}
Given the argument above, since $n$ has a maximum value, we have a maximum value for the momentum eigenvalue as well.
In turn, this implies a maximum value for the energy of a particle in a box.
This is depicted in Fig.\ref{fig:box} for the case $L=1$ m and $L=20\ell_\text{Pl}$.

It is worth noticing that no further constraint is present.
In particular, the size of the box, or any other length scale, is not quantized in this framework.
Rather, a minimum wavelength for the particle in the box is present, corresponding to the maximum momentum allowed.
It is of the order of the Planck length, consistently with the assumption of a minimal uncertainty in position.
This implies a minimal size for a non-empty box.
Such aspects may have interesting implications in some physical scenarios, such as the case of Casimir effect between two parallel plates.
In fact, such a system behaves as a box, with the plates effectively acting as walls.
Thus, reducing the spacing between the plates, on one hand the number of allowed states decreases.
On the other hand, the energy associated with each standing wave increases, since it depends on the inverse of the box size.
However, larger energies become less and less probable, producing an overall reduction of the total energy between the plates.
Such reduction, though, is expected to be faster than in the standard case due to the decreased number of allowed states.
Finally, when the threshold in Eq.\eqref{eqn:threshold} is crossed, no energy state is allowed in the box, producing a diverging attractive force between the plates.
Similar results have been obtained heuristically in \cite{Blasone:2019wad}.

The case of a vanishing $f(p)$ deserves a special remark.
In fact, according to Eq.\eqref{eqn:limited_p}, especially the case $|\delta_0|< \sqrt{-\epsilon_0}$, $p_0$ is not necessarily bounded.
Thus, by Eq.\eqref{eqn:wavenumber_box}, any integer $n$ and any length $L$ are allowed.
However, at the same time the quantity $p$ is bounded.
We thus have the alternative result in which an infinite number of standing waves is still possible, no forcibly empty boxes are predicted, but the system still exhibits a finite maximum momentum $p$, this time understood as a limit value when $n\rightarrow\infty$.


\section{Potential barrier} \label{sec:barrier}

As a further analysis, let us consider the case of a rectangular potential barrier of height $U_0$ and width $\Delta$.
Specifically, using the same argument applied in the previous section to determine the quasi-position coordinate of the box's walls, we will assume that the barrier is between the values $\xi = 0$ and $\xi = \Delta$.
Thus, using Eq.\eqref{eqn:momentum_eigen_position}, for $\xi < 0$ we have a linear combination of an incoming and a reflected waves
\begin{equation}
	\psi_1(\xi) = \frac{1}{\sqrt{2 \pi \hbar}} \left\{\mathcal{A}_1 \frac{\chi(p)}{f(p)} \exp \left[i \frac{\xi p_0(p)}{\hbar}\right] + \mathcal{B}_1 \frac{\chi(-p)}{f(-p)} \exp \left[i \frac{\xi p_0(- p)}{\hbar}\right]\right\},
\end{equation}
where $p$ is the momentum of the free particle.
For $\xi > \Delta$, we have an outgoing, transmitted wave
\begin{equation}
	\psi_3(\xi) = \frac{\mathcal{A}_3}{\sqrt{2 \pi \hbar}} \frac{\chi(p)}{f(p)} \exp \left[i \frac{\xi p_0(p)}{\hbar}\right].
\end{equation}
As for the region $0 < \xi < \Delta$, we have a momentum
\begin{equation}
p' = \sqrt{2 m (E - U_0)}.
\end{equation}

When $E \geq U_0$, $p'$ is a real quantity.
Furthermore, $p' < p$ and, since the function $p_0(p)$ is a monotone, growing function, $p_0(p') < p_0(p)$.
Thus, on the two sides of the barrier we find right-moving waves of wave number $k_\text{R} = p_0(p)/\hbar$ and a left-moving wave of wave number $k_\text{L} = -p_0(-p)/\hbar$, while through the barrier we have a right-moving wave of wave number $k_{\text{R}}' = p_0(p')/\hbar$ and a left-moving wave of wave number $k_{\text{L}}' = -p_0(-p')/\hbar$.
Specifically, the wave function through the barrier is
\begin{equation}
	\psi_2(\xi)
	= \frac{1}{\sqrt{2 \pi \hbar}} \left\{\mathcal{A}_2 \frac{\chi(p')}{f(p')} \exp \left[i \xi k_{\text{R}}'\right] + \mathcal{B}_2 \frac{\chi(-p')}{f(-p')} \exp \left[- i \xi k_{\text{L}}'\right]\right\}.
\end{equation}
Using the relevant boundary condition for the present problem, we then find the following ratios
\begin{align}
	\frac{\mathcal{B}_1}{\mathcal{A}_1} = & - \frac{(k_\text{L}' + k_\text{r}) (k_\text{R}' - k_\text{R}) \left[e^{i \Delta  (k_\text{L}' + k_\text{R}')} - 1\right]}{(k_\text{L}' - k_\text{L}) (k_\text{R}' - k_\text{R}) e^{i \Delta  (k_\text{L}' + k_\text{R}')} - (k_\text{L}' + k_\text{R}) (k_\text{R}' + k_\text{L})} \frac{\chi(p)}{\chi(-p)} \frac{f(-p)}{f(p)}, \\
	\frac{\mathcal{A}_3}{\mathcal{A}_1} = & - \frac{(k_\text{L}' + k_\text{R}') (k_\text{L} + k_\text{R}) e^{i \Delta  (k_\text{R}' - k_\text{R})}}{(k_\text{L}' - k_\text{L}) (k_\text{R}' - k_\text{R}) e^{i \Delta  (k_\text{L}' + k_\text{R}')} - (k_\text{L}' + k_\text{R}) (k_\text{R}' + k_\text{L})}.
\end{align}
Such ratios, the absolute values of which corresponds to the reflection and transmission coefficients, respectively, acquire the same form to those in the standard theory for model with an even $f(p)$.
In fact, in such a case we have $k_\text{R} = k_\text{L}$ and $k_\text{R}' = k_\text{L}'$.
Similar conclusions are obtained for the case $E < U_0$, with the obvious difference that in such a case $k_\text{R}', k_\text{L}' \in \mathbb{C}$, although it does not need to be purely imaginary.

\begin{figure}
	\centering
	\includegraphics[width=0.8\textwidth]{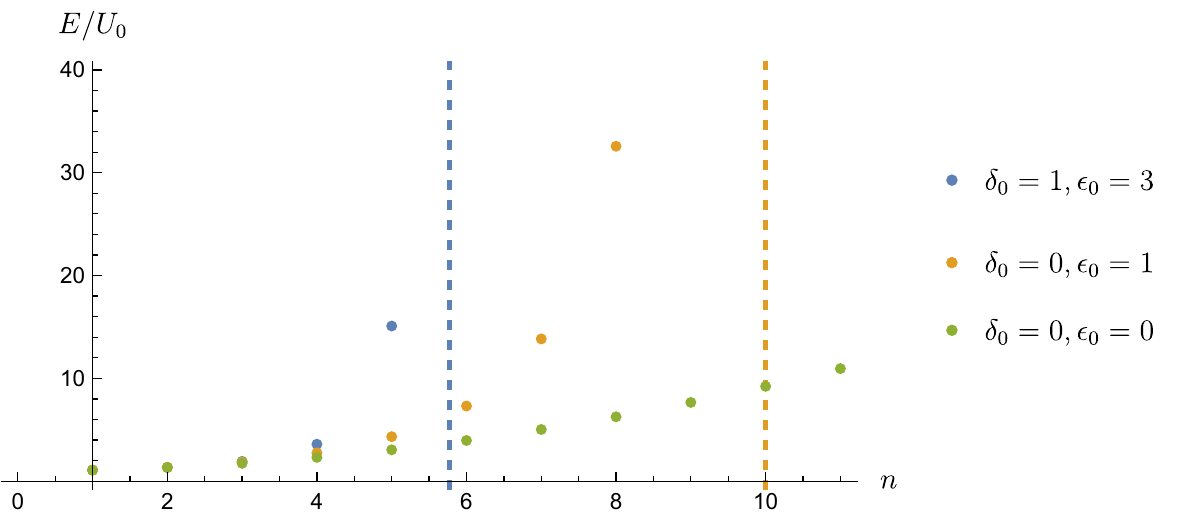}
	\caption{Values of $E/U_0$ corresponding to transmission resonances for $\Delta = 20 \ell_\text{Pl}$ and $U_0 = 60 \hbar^2 / m \Delta^2$.
		The dashed vertical lines correspond to the maximum number of resonances present in that particular model.}
	\label{fig:resonances}
\end{figure}

In terms of the momenta $p$ and $p'$, we have new effects due to GUP.
For practical purposes, let us consider the model in Eq.\eqref{eqn:GUP}.
Then,
\begin{equation}
	k_{\text{R,L}} = \frac{p_0}{\hbar} = \frac{1}{\hbar \sqrt{\epsilon}} \arctan\left[\frac{- \delta \pm (\delta^2 + \epsilon) p}{\sqrt{\epsilon}}\right] + \frac{1}{\hbar \sqrt{\epsilon}} \arctan\left(\frac{\delta}{\sqrt{\epsilon}}\right),
\end{equation}
\begin{figure}
	\centering
	\begin{subfigure}{0.48\textwidth}
		\includegraphics[width=\textwidth]{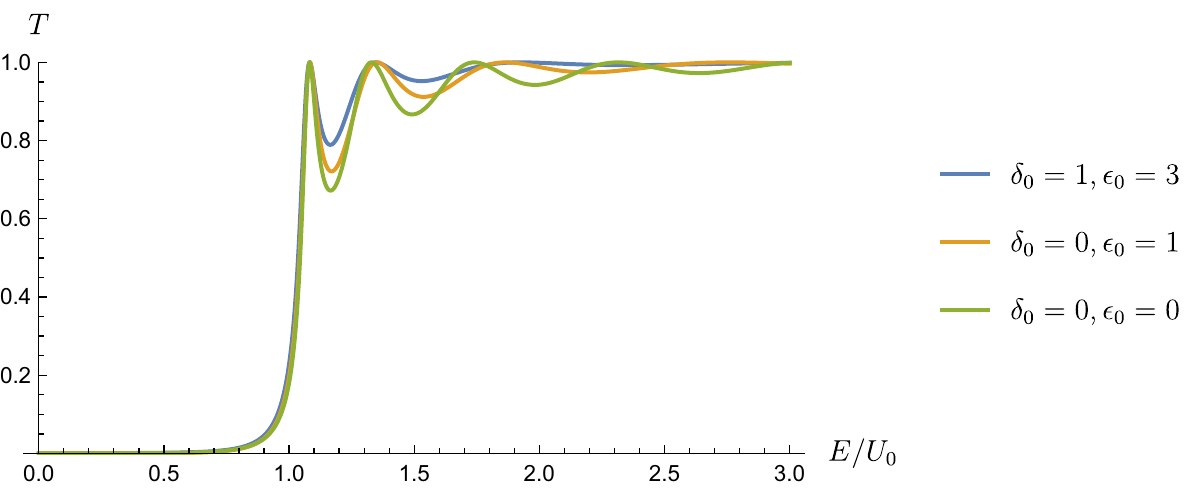}
		\caption{}
		\label{fig:transmission_larger}
	\end{subfigure}
	\hfill
	\begin{subfigure}{0.48\textwidth}
		\includegraphics[width=\textwidth]{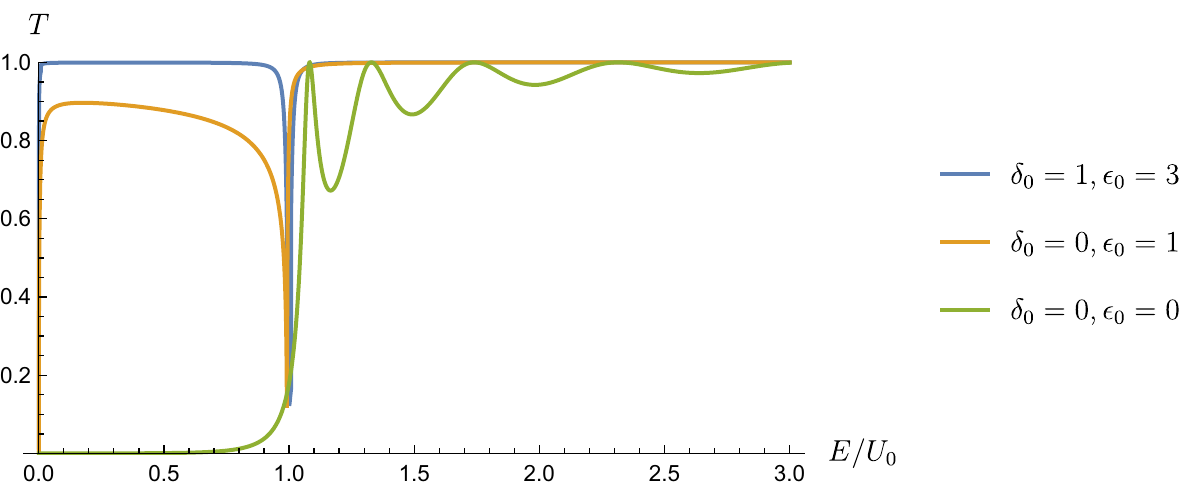}
		\caption{}
		\label{fig:transmission_smaller}
	\end{subfigure}
	\caption{Transmission coefficient $T = |\mathcal{A}_3/\mathcal{A}_1|$ as a function of $E/U_0$.
		Here, we considered a potential barrier with $\Delta = 20 \ell_\text{Pl}$ (Fig.\ref{fig:transmission_larger}) and one with $\Delta = 1 \ell_\text{Pl}$ (Fig.\ref{fig:transmission_smaller}).
		In both cases, we considered $U_0 = 60 \hbar^2/m\Delta^2$.}
\end{figure}
and similarly for $k_\text{R,L}'$ in terms of $p'$.
Then, as it can be seen from Fig.\ref{fig:transmission_larger}, the profile for the transmission coefficient $T=|\mathcal{A}_3/\mathcal{A}_1|$ is different from the one in the standard theory, in green.
Specifically, it is worth noticing that the resonances are at different values of energy with respect to the standard case.
In fact, maxima for the transition coefficient for $E\geq U_0$ are for $(k_{\text{R}}' + k_{\text{L}}') \Delta = 2 n \pi$, with $n$ a positive integer.
Thus, maxima are for values of energy such that
\begin{equation}
	\frac{E}{U_0} = \frac{1}{2 m U_0} \left[\frac{\sqrt{\delta^2 \sin^2 \left(\frac{2 \pi n \hbar \sqrt{\epsilon}}{\Delta}\right) + \epsilon} + \sqrt{\epsilon} \cos \left(\frac{2 \pi n \hbar \sqrt{\epsilon}}{\Delta}\right)}{\left(\delta^2 + \epsilon\right) \sin \left(\frac{2 \pi n \hbar \sqrt{\epsilon}}{\Delta}\right)}\right]^2 + 1
\end{equation}
Such relation is shown in Fig.\ref{fig:resonances} for the model in Eq.\eqref{eqn:GUP} and is compared with the standard model.
It is worth observing that, differently from the standard theory, GUP introduces a finite number of resonances.
The maximum number is given by
\begin{equation}
	n_\text{max} = \left\lfloor\frac{\Delta}{2 \sqrt{\epsilon} \hbar}\right\rfloor .
\end{equation}
In this case as well, such maximum number can be explained in terms of a minimal length.
Furthermore, this last relation predicts cases in which no resonance is present.
When this happens, that is, when
\begin{equation}
	\Delta < 2 \ell_{\text{Pl}} \sqrt{\epsilon_0},
\end{equation}
one finds that the transmission coefficient is appreciably different from zero even when $E < U_0$, as we can see from Fig.\ref{fig:transmission_smaller} for the case $\Delta = 1 \ell_{\text{Pl}}$.
This is consistent with a minimal uncertainty in position.
In fact, when the width of the barrier is smaller or compatible with the minimal uncertainty introduced by GUP, $\Delta q_{\text{min}} = \ell_{\text{Pl}} \sqrt{\epsilon_0}$ for the model in Eq.\eqref{eqn:GUP}, the barrier is not efficient in blocking the incoming wave.

\section{Conclusions} \label{sec:conclusions}

In this paper, we have analyzed a generic modification of QM presenting a minimal length as derived by the commutation relation Eq.\eqref{eqn:GUP_gen}.
Furthermore, we specialized the results for a particular function $f(\hat{p})$ consisting in a second degree polynomial.
In both cases, we focused on the properties of the function $f(\hat{p})$ defining the modification, on the construction of maximally localized states compatible with the models, and the integral transforms to the corresponding quasi-position space.
In particular, we found the representation of the position and momentum operators in this new space.
Moreover, we analysed the auxiliary quantity $p_0$ corresponding to the conjugate momentum to the position $q$.
We found that the momenta $p$ and $p_0$ have effectively equivalent roles in describing momentum.
Finally, we considered the case of a particle in a box to study the properties of such a system under the influence of a minimal length.

As for the model in Eq.\eqref{eqn:GUP}, although following the analysis in \cite{Kempf1995_1}, the presence of a linear term in $\hat{p}$ in Eq.\eqref{eqn:GUP} produces notable effects, absent in \cite{Kempf1995_1}.
One interesting formal aspect, shared with the model in \cite{Kempf1995_1}, is that the quantity $p_0$ acquires values on a limited interval.
However, by imposing the condition $p_0 \simeq p$ for small values of $p$, the same interval is not symmetric about $p=0$.
This is an effect of the anisotropic character of this particular model.
We concluded the analysis of this model by showing approximated relations concerning the position and momentum operators.

As for the example of a particle in a box, we found that a minimal size is present for the box to not be empty.
Furthermore, in general, a finite number of energy eigenstates are allowed in the box, depending on the size of the box.
This is in strong contrast with the standard result in which an infinite number of states are always available, regardless of the size of the box.
It is interesting to notice that the minimal size of a non-empty box is of the order of the minimal length $\ell$ and the number of allowed states is of the order of the ratio $L/\ell$ of the size of the box and the minimal length.
Similarly, in the case of the potential barrier we found that only a finite number of resonances are allowed, in contrast with the infinite number of resonances predicted by the standard theory.
In this case as well, the maximum number of resonances is of the order of the ratio between the barrier's width and the minimal length allowed in the model.

Finally, it is worth noticing that, in case the choices of the parameters for the second-degree model allow $f(p)<0$, since we are forced to restrict the domain of the physical momentum $p$, we have interesting deviations from the behavior shown above.
In particular, when $|\delta_0| < \sqrt{-\epsilon_0}$, with $\epsilon_0 < 0$, no bound is present for the auxiliary momentum $p_0$.
This signifies that we would get an infinite number of states for the particle in the box and an infinite number of resonances in the case of the potential barrier, similar to standard QM.
However, such energy states and such resonances are all characterized by finite energies.
In fact, being the physical momentum $p$ restricted to a finite interval, the energy is finite as well.

This analysis serves to shed some light on common misconceptions present in the literature regarding models of QM with a minimal length.
In particular, it is clear that, as already shown in \cite{Kempf1995_1}, a position representation similar to that of QM is not possible and the best one can do is resorting to the momentum and quasi-position representations.
Furthermore, the scalar product of two functions in quasi-position representation is not given by the usual relation of standard QM, but involves the transform of the cross-correlation of the two functions.
As we have seen, this is the result of an invertible integral transform between momentum and quasi-position spaces and the side-effect of non-orthogonality of maximally localized functions.
In fact, the same applies to standard QM when one uses minimal uncertainty product functions, \emph{i.e.} Gaussian functions.
Finally, the position operator in quasi-position space is not a multiplicative operator.
Rather, the usual multiplicative term is accompanied by the momentum operator multiplied by an imaginary quantity and, possibly, by an additive imaginary constant.
In fact, compatibly with the integral transform between momentum and quasi-position spaces, the multiplicative term is not Hermitian and the momentum operator and the corresponding imaginary coefficient are necessary to make the position operator Hermitian.

As a final remark, it is worth highlighting that here we presented the case of a one-dimensional, non-relativistic system.
It is known that accounting for a higher number of dimensions implies a richer structure for the model.
In particular, one has to account for the necessary non-commutativity of position \cite{Kempf1995_1,Mignemi2012,Pramanik2013_1}.
Thus, an analysis in such extended framework is expected to show a likewise richer structure for the position operator.
We will present and comment on such aspects in a future publication.

\section*{Acknowledgments}

The author would like to thank the anonymous Referees for their comments, which greatly improved the content of the manuscript.


\end{document}